\documentclass[11pt,a4paper]{article}
\pdfoutput=1
\usepackage{jheppub}
\usepackage{amsfonts, amssymb, amsmath}
\usepackage{mathrsfs}
\usepackage{graphicx}
\usepackage[utf8]{inputenc}
\usepackage[russian,english]{babel}

\usepackage{color}
\definecolor{dark-gray}{gray}{0.20}
\definecolor{gray}{gray}{0.30}
\definecolor{light-gray}{gray}{0.80}
\definecolor{dark-red}{rgb}{0.7,0,0}
\definecolor{dark-green}{rgb}{0.1,0.4,0}
\definecolor{dark-blue}{rgb}{0.3,0.3,0.7}
\definecolor{light-blue}{rgb}{0.8,0.8,1}
\definecolor{blue}{rgb}{0,0,1}
\definecolor{red}{rgb}{1,0,0}
\definecolor{green}{rgb}{0,1,0}

\usepackage{hyperref}
\hypersetup{
	colorlinks=true,
	linkcolor=dark-blue,
	citecolor=dark-red,
	urlcolor=dark-blue,
	linktoc=page
}

\def\cF{{\cal F}}

\def\cN{{\cal N}}

\def\U{{\rm U}}
\def\SU{{\rm SU}}

\newcommand{\be}{\begin{equation}}
\newcommand{\ee}{\end{equation}}
\newcommand{\bea}{\begin{eqnarray}}
\newcommand{\eea}{\end{eqnarray}}

\title{The dark (BPS) side of\\ thermodynamics in Minkowski$_4$}

\author{Kiril Hristov}

\affiliation{Faculty of Physics, Sofia University, J. Bourchier Blvd. 5, 1164 Sofia, Bulgaria}
\affiliation{INRNE, Bulgarian Academy of Sciences, Tsarigradsko Chaussee 72, 1784 Sofia, Bulgaria}

\emailAdd{khristov@phys.uni-sofia.bg}

\abstract{
\noindent We explore the BPS limit of asymptotically flat black hole thermodynamics, drawing analogies with recent studies of black holes in anti-de Sitter space. Although ultimately motivated by supersymmetry and quantum gravity, the BPS limit is classically well-defined for the dyonic Kerr-Newman black holes in pure Einstein-Maxwell theory and various generalizations with additional scalars and gauge fields that admit embedding in $\cN=2$ supergravity (for concreteness we consider the STU model). This limit is manifestly different from extremality as it includes a family of {\it rotating} Euclidean saddles that represent a (fictitious) thermal branch, smoothly connected with the extremal static Reissner-Nordstr\"om black hole and its generalizations with extra matter. We are able to evaluate unambiguously all BPS chemical potentials, finding constant imaginary angular velocity as a consequence of Wick rotation. In the mixed ensemble of fixed magnetic and varying electric charges, we derive the OSV formula that is crucial for the microscopic understanding of the system, and provide evidence that it should be interpreted as a fixed-point formula.
}
\date{\today}
\begin{document}
\maketitle


\section{Introduction and main results}
\label{sec:intro}
Black hole thermodynamics is by now an old and well-explored subject that started around 50 years ago,~\footnote{The album that inspired the present title came out at the same time.} most notably fueled by the works of Bekenstein \cite{Bekenstein:1973ur} and Hawking \cite{Hawking:1975vcx}. Yet to this day the underlying microscopic degrees of freedom remain one of the great open questions and consequently a major source of inspiration for innumerable novel ideas in the area of quantum gravity. One of the successes of string theory as a proposed theory of quantum gravity is the microscopic entropy counting of a supersymmetric class of asymptotically flat extremal Reissner-Nordstr\"om black holes in theories of supergravity with multiple scalars and abelian gauge fields, see \cite{Strominger:1996sh} and innumerable references thereof.

A more recent development, fueled by the AdS/CFT correspondence \cite{Maldacena:1997re}, is the microscopic entropy counting of supersymmetric black holes in anti-de Sitter (AdS) space starting with \cite{Benini:2015eyy}, see \cite{Zaffaroni:2019dhb} for a review. One of the main advances from the study of AdS black holes is the detailed understanding of the relation between the (generally complex) on-shell action and the black hole entropy, as well as the discovery of new Euclidean backgrounds that should be included in the path integral along with the black holes. In this note we borrow some of these ideas and come back to the asymptotically flat case. Even though black holes in Minkowski were chronologically considered and understood much earlier, we are still able to report a number of new and perhaps not entirely expected results.

More precisely, we focus on the set of solutions that obey the so-called BPS limit, and describe their thermodynamics. By a BPS limit we mean the condition that supersymmetry is preserved, whenever we start with a theory of supergravity. Since this condition is in practice a simple restriction on the parameter space of purely bosonic solutions, such a BPS limit is well-defined also in theories where the fermion sector is different or altogether absent. The simplest example is pure Einstein-Maxwell theory, which admits embedding in minimal $\cN=2$ supergravity. Since the set of bosonic solutions coincide, we simply define the BPS limit for the black holes as inherited from the larger theory. Our analysis here is only semi-classical and none of the calculations and results depend explicitly on the fermionic sector. We caution the reader that a full quantum gravity calculation will likely change drastically this conclusion, as suggested by recent results in \cite{Iliesiu:2020qvm,Heydeman:2020hhw}, in which case including the corresponding fermionic sector will ensure the present results remain valid.

\begin{figure}[h]
\label{fig}
\includegraphics[width=13cm]{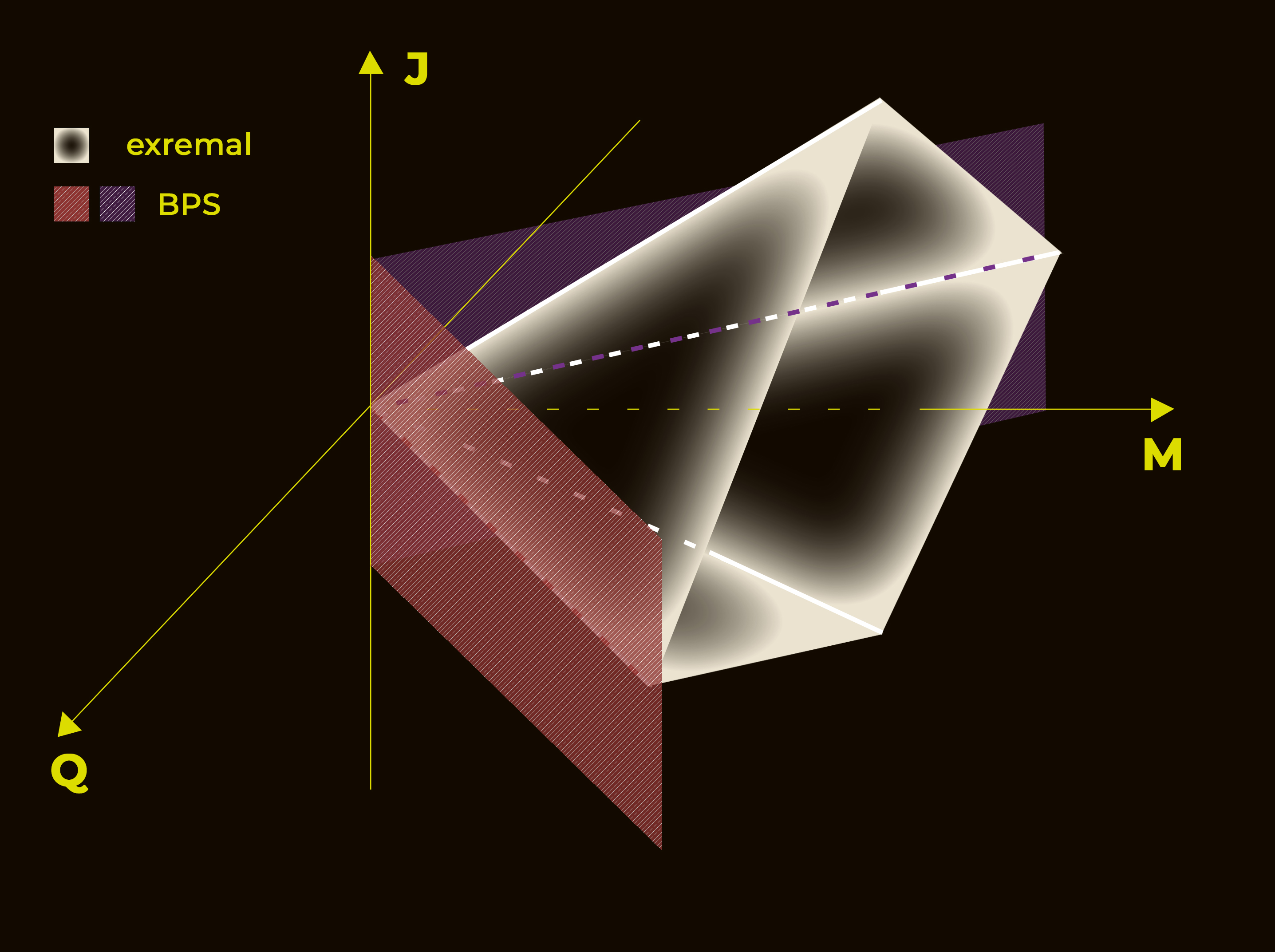}
\centering
\caption{The black hole solution space in the case where only mass ($M$), angular momentum ($J$), and a single electric charge ($Q$), are allowed. The sides of the pyramid are the extremal black holes, while the inside is populated by the conventional thermal solutions (the base extends to $M \rightarrow \infty$). The BPS family of solutions is depicted by the two colored vertical planes (which also extend to $J \rightarrow \pm \infty$) that only intersect the two edges of the pyramid on the $J = 0$ plane. Note that in reality the space of thermal black hole solutions does not exhibit sharp edges and is cigar rather than pyramid shaped, but the depiction emphasizes more clearly the crucial distinction between the extremal and the BPS surfaces and their intersection on the two lines.}
\end{figure}

It is important to stress here that the BPS limit does {\it not} coincide with the extremal limit~\footnote{In this sense the present construction is different from Sen's entropy function, \cite{Sen:2005wa}, that relies on the existence of a near-horizon region at extremality.} of vanishing temperature for the black holes in Einstein-Maxwell theory (Fig.\ \ref{fig}) or the more general STU model we consider, found in \cite{Chow:2013tia,Chow:2014cca}. Both limits define a codimension 1 hypersurface on the full parameter space, and their intersection is a codimension 2 hypersurface of supersymmetric extremal black holes. The widespread terminology in the literature is that {\it supersymmetric black holes} denotes only the codimension 2 intersection between the BPS and extremal limits, due to the fact that the rest of the solutions in the BPS limit do not have an interpretation as Lorentzian black holes. Upon Wick rotation to Euclidean signature, we show that the BPS limit produces regular solutions without further restriction except the periodic identification of the (former) time coordinate, in analogy to the Euclidean saddles with AdS asymptotics in \cite{BenettiGenolini:2019jdz,Bobev:2020pjk}. We therefore define the {\it fictitious temperature}, whose inverse we denote here by $\beta_\tau$ in order to distinguish with the inverse temperature $\beta$ of the Lorentzian black holes. It was shown in \cite{Cabo-Bizet:2018ehj} and \cite{Cassani:2019mms} for one class of black holes in AdS that one can define the BPS thermodynamics with $\beta_\tau$ as a particular BPS limit of the conventional black holes thermodynamics with $\beta$. Largely following these two references, we find very similar results for black holes in Minkowski$_4$.

We can briefly summarize our findings as follows. For conventional thermodynamics with $\beta$, the on-shell action $I$ satisfies the {\it quantum statistical relation}, \cite{Gibbons:1976ue,Gibbons:2004ai}
\be
\label{eq:convth}
	I = \beta\, M - S - \beta\, \Omega\, J - \beta\, \Phi^I\, Q_I - \beta\, \Psi_I\, P^I\ ,
\ee
for asymptotically flat black holes of mass $M$, angular velocity and momentum $\Omega$ and $J$, a set of electric chemical potentials and conjugate charges $\{ \Phi^I, Q_I\}$, and a set of magnetic chemical potentials and conjugate charges $\{ \Psi_I, P^I \}$. The first law of thermodynamics in this grand-canonical ensemble of fixed $\beta, \Omega, \Phi^I, \Psi_I$ simply states that $\delta I = 0$ such that $I = I (\beta, \Omega, \Phi^I, \Psi_I)$. The direct extremal limit, $\beta \rightarrow \infty$, is not well-defined since most quantities on the r.h.s.\ of \eqref{eq:convth} diverge.

In the BPS limit, which gives a quadratic constraint on the black hole mass in terms of the electromagnetic charges (see \eqref{eq:BPScont} and \eqref{eq:stubpscont}), we define the following variables that play the role of BPS chemical potentials, as in \cite{Cabo-Bizet:2018ehj,Cassani:2019mms},
\be
	\omega := \beta_\tau\, (\Omega - \Omega^*)\ , \quad \varphi^I := \beta_\tau\, (\Phi^I - \Phi^{I, *})\ , \quad \xi_I := \beta_\tau\, (\Psi_I - \Psi_I^*)\ ,
\ee
where $*$ denotes the {\it extremal} value of the corresponding BPS quantity. The BPS value of the angular momentum $J$ is arbitrary while imposing extremality in addition sets $J^* = 0$, since the codimension 2 intersection of the BPS and extremal surfaces are the static zero-temperature Reissner-Nordstr\"om black holes (see Fig.\ \ref{fig}) and generalizations with extra matter. The BPS thermodynamics follows from the BPS quantum statistical relation
\be
	I = -S - \omega\, J- \varphi^I\, Q_I - \xi_I\, P^I\ ,
\ee
with the first law of BPS thermodynamics $\delta I = 0$, such that $I = I (\omega, \varphi^I, \xi_I)$. Crucially, the BPS on-shell action is shown to be {\it independent} of the fictitious temperature $\beta_\tau$, allowing us a free passage to the extremal limit, $\beta_\tau \rightarrow \infty$. Another important result is that the BPS angular velocity is always fixed to a particular imaginary value,
\be
	\omega = \pm 2\, \pi\, {\rm i}\ ,
\ee
on the two BPS branches (defined in the main text). It follows that one is forced to stay in the grand-canonical ensemble of fixed $\omega$ and varying angular momentum $J$, while freely changing to the microcanonical ensemble fixing some of the electromagnetic charges $Q_I, P^I$. It is in the mixed ensemble of fixed magnetic and fluctuating electric charges where the (Legendre transformed) on-shell action $\tilde I (\omega, \varphi^I, P^I)$ takes the simple and suggestive form
\be
	\tilde I = \pm \frac{{\rm i}\, \pi^2}{\omega}\, {\rm Im} F(4\, P^I + \frac{{\rm i}}{2 \pi}\, \varphi^I)\ ,
\ee
where $F$ is the supergravity prepotential that uniquely defines the Lagrangian of the theory, for the STU model given in \eqref{eq:sqrt}, for minimal supergravity (or Einstein-Maxwell theory) in \eqref{eq:minprep}. Thus we recover the OSV formula proposed in \cite{Ooguri:2004zv} in order to calculate the exact quantum entropy of the extremal BPS black holes. The same on-shell action is now shown to hold for the full set of BPS solutions, but the direct relation with the entropy is only valid upon imposing extremality, due to the extra identity
\be
	S^* = S + \omega\, J\ ,
\ee
ensuring that
\be
	\tilde I = - S^* - \varphi^I\, Q_I\ .
\ee

The rest of this note is organized in a way that allows readers uninterested in supersymmetry to follow easily all calculations and implications, as follows. In section \ref{sec:setup} we briefly introduce the gravitational theory we consider, the general black hole solutions and their conventional thermodynamics. In section \ref{sec:flat} we focus on the solutions in Einstein-Maxwell theory, defining the BPS limit and elaborating on the resulting BPS thermodynamics in section \ref{subsec:bpsflat} and demonstrating the regularity of all solutions in Euclidean signature in section \ref{subsec:wick}. The rest of the paper requires more specific supesymmetry knowledge and interest - in section \ref{sec:AdS} we consider the general STU model, giving more details about the underlying superalgebra that ultimately defines the BPS limit. In section \ref{subsec:qsrstu} we discuss the BPS thermodynamics for the black holes in the STU model, and in section \ref{subsec:osv} we discuss the relation of our results with the OSV formula and the more recent proposal of gravitational block gluing, \cite{Hosseini:2019iad,Hristov:2021qsw}. We finish in section \ref{sec:outro} with further discussion and open questions.

\section{Black holes and conventional thermodynamics, a sketch}
\label{sec:setup}
We first need to write down and characterize the basic properties of the black holes in the STU model and its truncation to Einstein-Maxwell theory. For simplicity and clarity of the presentation, we are going to follow strictly the conventions and notations of \cite{Chow:2013tia,Chow:2014cca},~\footnote{We explicitly note below each case of notational change.} which wrote down the general {\it seed} solution~\footnote{Due to the lack of scalar potential in theories exhibiting Minkowski vacua, the scalars can take arbitrary asymptotic values in all black hole solutions. The seed solution refers to the fact that the scalars are fixed to particularly chosen constants for simplicity, without any restrictions on the conserved asymptotic charges of the black hole. The most general case with arbitrary asymptotic scalars can be generated by additional dualities (e.g.\ in string theory), but it does not concern directly our analysis here.} for the thermal black holes.  The full Lagrangian and all details of the black hole solutions are carefully recorded in the original references, \cite{Chow:2013tia,Chow:2014cca}, building on numerous previous works including \cite{Rasheed:1995zv,Cvetic:1996kv,Behrndt:1996hu,Larsen:1999pp,Lozano-Tellechea:1999lwm,Bertolini:2000ei,Chong:2004na,Bellucci:2008sv,DallAgata:2010srl} and references therein and thereof. We are only going to sketch some aspects that need to be highlighted in relation to the present scope, starting with the theory under consideration and moving to the form of the metric and the general thermodynamic properties.

\subsection{Main aspects of the theory}
The supergravity theory that goes under the name {\it STU model} is an $\cN=2$ supergravity coupled to 3 vector multiplets, see \cite{Andrianopoli:1996cm} for standard supergravity quantities and conventions. The supergravity multiplet consists of the metric, a gauge field and two gravitini, while each vector multiplet contains one gauge field, a complex scalar and a spin $1/2$ fermion. Together, the bosonic content is therefore the metric, four $\U(1)$ fields $A^I$, $I=0, 1, 2, 3$,~\footnote{In \cite{Chow:2013tia,Chow:2014cca} the authors choose to label the gauge fields with $I=1, 2, 3, 4$ but we stick to the standard supergravity conventions \cite{Andrianopoli:1996cm}.} and three complex scalars $z^i = e^{- \varphi_i} + {\rm i} \chi_i$, $i = 1, 2, 3$. These scalars parametrize the coset space $[\SU(1,1)/\U(1)]^3$, and are sometimes denoted S, T and U (giving the name of the model). This model can be further embedded in thr maximal $\cN=8$ supergravity in 4d, and the Lagrangian can be formulated in a number of different electro-magnetic frames making use either of the original gauge fields or their magnetic duals, see \cite{Chow:2014cca} for all details.

An important point about $\cN=2$ supergravity of relevance here is that the Lagrangian (in a certain electro-magnetic frame) is uniquely specified by the so-called {\it prepotential} $F (X)$, a holomorphic function of projective coordinates $X^I$ that determine the complex scalars, $z^i := X^i / X^0$. In the frame in which the black hole solutions \cite{Chow:2013tia,Chow:2014cca} are explicitly presented, the corresponding prepotential has a square root form
\be
\label{eq:sqrt}
	F_\text{STU} (X) =- 2 {\rm i}\, \sqrt{X^0 X^1 X^2 X^3}\ ,
\ee
which can be also brought to the cubic form $F = - X^1 X^2 X^3 / X^0$ after a suitable symplectic rotation that essentially exchanges one electric charge for a magnetic one, and vice versa, see \cite{Chow:2014cca}. Choosing to write the Lagrangian in the square root prepotential frame \eqref{eq:sqrt} has the advantage that truncating the theory to smaller subsectors is rather straightforward, as again explained in detail in the original reference. One possible supersymmetry-preserving truncation is going back to minimal supergravity with no extra vector multiplets, reached by simply identifying all gauge fields $A^0 = A^1 = A^2 = A^3 = A$ and setting $X^0 = X^1 = X^2 = X^3 = X$ such that all scalars are truncated away. The resulting prepotential is simply
\be
\label{eq:minprep}
	F_\text{min} (X) = - 2 {\rm i}\, (X)^2\ ,
\ee  
and the bosonic part of the theory is pure Einstein-Maxwell theory:
\be
\label{eq:minlag}
	{\cal L} = R * 1 - 2 *F \wedge F\ ,
\ee
with $F = {\rm d} A$. Another prominent truncation is the so-called $X^0 X^1$ model obtained by a pairwise identification e.g.\ $X^0 = X^2, X^1 = X^3$, keeping one complex scalar. Note that the entire STU model, as well as its truncations, can also be embedded in higher-dimensional supergravity and string theory, such that we can ensure a full UV completion. This will not play a role here as we focus only on semi-classical analysis, but is important in view of the relation with the OSV formula discussed further in section \ref{subsec:osv}.

\subsection{Main aspects of the solutions}
The black holes of \cite{Chow:2013tia,Chow:2014cca} in the full STU model depend in total on 11 free parameters that in turn (see below) determine the set of conserved asymptotic charges: a mass parameter $m$, NUT parameter $n$, rotation parameter $a$, electric charge parameters $\delta_I$ and magnetic charge parameters $\gamma_I$. Standard black hole thermodynamics is defined for a vanishing NUT charge, fixing the parameter $n$ in terms of the other parameters. As discussed above, the scalars have been fixed to particular constants at infinity, corresponding to $\varphi_i (\infty) = \chi_i (\infty) = 0$, $\forall i$. 

The solutions are carefully spelled out in the original references and here we can illustrate the main features by simply considering the limit to the Einstein-Maxwell theory, corresponding to $\delta_0 = \delta_1 = \delta_2 = \delta_3 = \delta$ and $\gamma_0= \gamma_1 = \gamma_2 = \gamma_3 = \gamma$.~\footnote{We remind the reader that in the original references, \cite{Chow:2013tia,Chow:2014cca}, the authors use index $4$ instead of $0$.} The metric is given by
\be
\label{eq:minmetric}
	{\rm d} s^2 = - \frac{R - U}{W} ({\rm d} t+\omega_3)^2 + W \left( \frac{{\rm d} r^2}{R} + \frac{{\rm d} u^2}{U} + \frac{R U}{a^2 (R-U)}\, {\rm d} \phi^2 \right)\ ,
\ee
with
\be
	R(r) = r^2 - 2 m r +a^2 - n^2\ , \qquad U(u) = a^2 - (u-n)^2\ ,
\ee
\be
	W^2 = (R - U)^2 + L^2 + 2 (R-U) (2 M r + V)\ , \qquad \omega_3 = \frac{U L}{a (R-U)}\, {\rm d} \phi\ ,
\ee
and 
\bea
\begin{split}
	L(r) &= \frac{1}{8 e^{4 \gamma} e^{4 \delta}}\, \Big[ 4 e^{2 (\gamma+\delta)} \left( (1+e^{4 \gamma}) (1+e^{4 \delta}) m + (1-e^{4 \gamma}) (1-e^{4 \delta}) n\right)\, r \\
&+ (m^2+n^2) \left( 1+12 e^{4 (\gamma+\delta)} +e^{8 \gamma}+ e^{8 \delta} + e^{8 (\gamma+\delta)}- 4 e^{2 (\gamma+\delta)} (1+e^{4 \gamma}) (1+e^{4 \delta})  \right)\Big]\ ,
\end{split}
\eea
with $V(u)$ a similar linear function, see \cite{Chow:2014cca}. Above, we already assumed that the total NUT charge is vanishing, $N=0$, fixing the parameter $n$ in a specific way in terms of $m, \delta, \gamma$, 
\be
\label{eq:vanNUT}
	n = \frac{(1-e^{4 \gamma}) (1-e^{4 \delta})}{(1+e^{4 \gamma}) (1+e^{4 \delta})}\, m\ ,
\ee
and the constant $M$ above is the physical mass, the conserved asymptotic charge from time translation invariance,~\footnote{We choose the convention $c = G_N = 1$ for simplicity, such that factors of $G_N$ do not appear in the Lagrangian \eqref{eq:minlag} or in any of the physical quantities.}
\be
	M = \frac{1+4 e^{4 (\gamma+\delta)} + e^{8 \gamma} +e^{8 \delta} + e^{8 (\gamma+\delta)}}{2 e^{2 (\gamma+\delta)} (1+e^{4 \gamma}) (1+e^{4 \delta})}\, m\ .
\ee
Apart from time translations, the spacetime has a manifest axial symmetry along the $\phi$-angle. The spherical part of the metric can be rewritten in the standard angular variables $\{\theta, \phi \}$ upon the coordinate change $u = n + a \cos \theta$, and the static limit is reached for $a \rightarrow 0$ since the angular momentum corresponds to
\be
	J = a M\ .
\ee
The solution for the gauge field $A$ can be compactly written as
\be
\label{eq:gaugefield}
	A = - W \frac{\partial}{\partial \delta} \left( \frac{{\rm d} t + \omega_3}{W} \right)\ ,
\ee
where the derivatives should be taken {\it before} setting $N=0$ via \eqref{eq:vanNUT}, leading to the conserved electric and magnetic charges
\be
	Q = \frac{(1+e^{8 \gamma}) (e^{4 \delta}-1)}{8 e^{2 \gamma} (1+e^{4 \delta})}\, m\ , \qquad P = \frac{e^{2 \delta} (e^{4 \gamma}-1)}{4 e^{2 \gamma} (1+e^{4 \delta})}\, m\ .
\ee
Note that in this parametrization the electric charge $Q$ vanishes upon $\delta \rightarrow 0$, while the magnetic charge $P$ vanishes upon $\gamma \rightarrow 0$.

The more general solutions of the STU model with an independent set of parameters $\delta_I$ and $\gamma_I$ corresponding to independent charges $Q_I$ and $P^I$ are given in a very similar manner, but with more involved explicit expressions for the conserved charges that can be found in \cite{Chow:2014cca}. Additionally, the three complex scalar fields exhibit non-trivial dependence on the radial and angular coordinates $r$ and $u$. They can be found explicitly in \cite{Chow:2014cca},~\footnote{\label{ft:2} Note a somewhat inconvenient choice of parametrization made in \cite{Chow:2014cca} - the scalars considered there $z^i = X^i/X^0$ parametrize the {\it cubic} STU model, i.e\ in terms of the sections in the square root form of the prepotential \eqref{eq:sqrt} we have
\be
	z^1 = \frac{\sqrt{X^0 X^1}}{\sqrt{X^2 X^3}}\ , \quad z^2 = \frac{\sqrt{X^0 X^2}}{\sqrt{X^1 X^3}}\ , \quad z^3 = \frac{\sqrt{X^0 X^3}}{\sqrt{X^1 X^2}}\ .
\ee
} and we comment on the relation with the attractor mechanism in section \ref{subsec:osv}.

\subsection{Conventional thermodynamics}
\label{subsec:thermo}
The black holes in general exhibit outer and inner horizons at $r = r_\pm$, the roots of the radial function $R(r)$. Although it makes sense to define the laws of thermodynamics on both horizons, here we are going to look only at the outer horizon assuming $r_+ \in \mathbb{R}$.~\footnote{Hence we are going to drop the index $+$ on the chemical potentials evaluated in \cite{Chow:2014cca}.} When the roots are instead complex (before Wick rotation), as in the next section, the ordering $r_- < r_+$ loses meaning and one considers both roots for BPS thermodynamics. Just as above, we are only going to focus on the minimal limit of the parameters, $\delta_0 = \delta_1 = \delta_2 = \delta_3 = \delta$ and $\gamma_0= \gamma_1 = \gamma_2 = \gamma_3 = \gamma$. The reader can find all the general definitions and explicit expressions for the conventional thermodynamic quantities in \cite{Chow:2014cca}.

In the minimal case, the horizons are situated at
\be
	r_\pm = m \pm \sqrt{M^2 - 16 (Q^2 + P^2) - a^2}\ ,
\ee
such that regular event  horizons shield the singularity only when
\be
\label{eq:extremality}
	M^2 \geq 16 (Q^2 + P^2) + a^2\ ,
\ee  
and the bound is saturated by the extremal black hole with double horizon and vanishing temperature.

All the thermodynamic quantities are then implicitly evaluated using a positive real value of $r_+$, such as the Bekenstein-Hawking entropy
\be
	S = \pi L(r_+)\ ,
\ee
the angular velocity (conjugate to the angular momentum)
\be
	\Omega = \frac{a}{L(r_+)}\ ,
\ee
and the temperature
\be
\label{eq:standardtemp}
	T = \frac{R' (r_+)}{4 \pi L(r_+)} = \frac{r_+ - m}{2 \pi L(r_+)}\ .
\ee
Likewise, the electric $\Phi^I$ and magnetic potentials $\Psi_I$ are evaluated from the knowledge of the gauge fields at the outer horizon. 

Given that the outer horizon is a positive real number (true for regular black holes, i.e.\ all points on and inside the pyramid on Fig.\ \ref{fig}), all the conserved charges and chemical potentials are real. They obey the first law of thermodynamics 
\be
	\delta M = T\, \delta S + \Omega\, \delta J + \Phi^I\, \delta Q_I + \Psi_I\, \delta P^I\ ,
\ee
where $\delta$ is an infinitesimal variation and should not be confused with the $\delta_I$ that parametrize the black hole solutions. It is clear that in the minimal limit all electric and magnetic potentials are equal (denoted by $\Phi$ and $\Psi$) and the first law takes the form
\be
	\delta M = T\, \delta S + \Omega\, \delta J + 4\, \Phi\, \delta Q +4\, \Psi\, \delta P\ .
\ee

We can then define the free energy in the {\it grand canonical} ensemble (fixed temperature, angular velocity and electromagnetic potentials) as
\be
	\cF := M - T\, S - \Omega\, J - \Phi^I\,  Q_I - \Psi_I\,  P^I\ .
\ee
while changes of ensemble require respective Lagendre transform of the free energy to the new thermodynamic potential of interest. One can altermatively divide by the temperature on both sides in order to obtain the {\it quantum statistical relation}, \cite{Gibbons:1976ue,Gibbons:2004ai}
\be
\label{eq:qsr}
	I = \beta\, \cF = \beta\, M - S - \beta\, \Omega\, J - \beta\, \Phi^I\, Q_I - \beta\, \Psi_I\, P^I\ ,
\ee
using the definition $\beta = 1/T$. The quantity on the left hand side is the {\it on-shell action} $I (\beta, \Omega, \Phi^I, \Psi_I)$ that can be computed independently by evaluating the classical action on the black hole solutions, having first Wick rotated them to Euclidean signature and identified periodically the time coordinate $\tau \sim \tau + \beta$.

Additionally, it can be checked that the asymptotically flat black holes satisfy the following Smarr relation,
\be
	M = 2\, T\, S + 2\, \Omega\, J + \Phi^I\, Q_I + \Psi_I\, P^I\ ,
\ee
which can be used freely to simplify the expression for the free energy or other thermodynamic quantities of interest, e.g.\ by direct substitution we find
\be
	\cF = T\, S + \Omega\, J\ .
\ee

\section{BPS thermodynamics in Einstein-Maxwell theory}
\label{sec:flat}
The BPS limit of the black hole solutions is dictated by the supersymmetry algebra, see more details in the next section when we discuss the full STU model. In the minimal theory, the BPS constraint is given by the following relation between the mass and electromagnetic charges, 
\be
\label{eq:BPScont}
	M^2 = 16 \left( Q^2 + P^2 \right)\ .
\ee
Note that the above relation does not feature the angular momentum $J$ and therefore allows it to be arbitrary and in particular {\it non-vanishing}, which sometimes mistakenly is claimed to also be a consequence of supersymmetry in the literature. This confusion has a simple explanation stemming from the wrongful identification of the BPS constraint \eqref{eq:BPScont} with the saturation of the extremality bound \eqref{eq:extremality}.~\footnote{Since this confusion is rather widespread, it will be unfair to attribute it to a particular reference. An example of such imprecise statements can be found in the author's PhD thesis, \cite{Hristov:2012bk}.} It is in fact easy to see that for $J \sim a \neq 0$, the BPS constraint violates \eqref{eq:extremality} and therefore produces a nakedly singular spacetime. This can be repaired by a Wick rotation to Euclidean signature, where these solutions can be regularized by a periodic identification of the (former) time direction, as we show in section \ref{subsec:wick}. Taking this for granted for the moment, we first describe the implications for the thermodynamics and define the new thermodynamic quantities featuring in the BPS quantum statistical relation, the BPS limit of \eqref{eq:qsr}.

\subsection{BPS quantum statistical relation}
\label{subsec:bpsflat}
As explained in \cite{Chow:2014cca}, the BPS constraint \eqref{eq:BPScont} can be efficiently reached by the following rescaling of the solution parameters,
\be
\label{eq:limitmin}
	m \rightarrow m\, \epsilon^2\ , \qquad e^\gamma \rightarrow \frac{e^\gamma}{\epsilon}\ ,
\ee
and then sending $\epsilon \rightarrow 0$, while keeping the rest of the parameters unchanged.~\footnote{\label{ft:1} There is an alternative way to reach the BPS constraint by interchanging the scaling of $\delta$ and $\gamma$ with $\epsilon$ that was not discussed in \cite{Chow:2014cca}. The choice in \eqref{eq:limitmin} allows for arbitrary magnetic charge, which however can never vanish (instead the electric charge is completely free). The opposite choice of exchanging $\delta$ and $\gamma$ fixes the magnetic charge to zero from the start with no other change on the remaining thermodynamic properties. Since we are interested in the generic case we only explicitly consider the limit \eqref{eq:limitmin}, stressing here that there is nothing special in switching off the magnetic charge.} Although not strictly needed for the limit and consequent analysis to make sense, we are also going to perform the change
\be
\label{eq:alimit}
	a \rightarrow {\rm i}\, a\ ,
\ee
based on the regularity of the solutions discussed in the next subsection.

We can now repeat the calculation for all thermodynamic quantities introduced in section \ref{subsec:thermo} in the limit \eqref{eq:limitmin}-\eqref{eq:alimit}, noting that we find
\be
\label{eq:bpshor}
	r_+ = \pm a\ ,
\ee
with both signs allowed.~\footnote{Alternatively one can choose to use $r_\pm = \pm a$ and look at BPS thermodynamics on both horizons instead of allowing both signs for $r_+$. The distinction is purely notational.} It is straightforward to evaluate the conserved charges and conjugate chemical potentials, noting that in the BPS limit defined above we recover {\it purely imaginary} expressions for the angular velocity and momentum,
\be
\label{eq:minbpspotentialsang}
	\Omega = \frac{4 {\rm i}\, a e^{4 \delta} (1+e^{4 \delta})^2}{e^{2 \gamma} (1+e^{8 \delta}) m \left(\pm 4 a e^{2\delta} (1+e^{4\delta}) + e^{2\gamma} (1+e^{8\delta}) m\right)}\ , \quad J = \frac{{\rm i}\, a e^{2\gamma} (1+e^{8\delta}) m}{2 e^{2 \delta} (1+e^{4 \delta})}\ ,
\ee
while all other quantities remain manifestly {\it real},
\bea
\begin{split}
\label{eq:minbpspotentials}
	\beta_\tau &=  \frac{\pi e^{2 \gamma} (1+e^{8 \delta}) m \left(4 a e^{2\delta} (1+e^{4\delta}) \pm e^{2\gamma} (1+e^{8\delta}) m\right)}{2 a e^{4 \delta} (1+e^{4 \delta})^2}\ , \quad M =  \frac{e^{2 \gamma} (1+e^{8 \delta}) m}{2 e^{2\delta} (1+e^{4\delta})}\ , \\
	\Phi &= \frac{(e^{8\delta} -1) \left(\pm 2 a e^{2\delta} (1+e^{4\delta}) + e^{2\gamma} (1+e^{8\delta}) m \right)}{(1+e^{8\delta}) \left(\pm 4 a e^{2\delta} (1+e^{4\delta}) + e^{2\gamma} (1+e^{8\delta}) m \right)}\ , \quad Q = \frac{e^{2\gamma} (e^{4\delta}-1) m}{8 e^{2\delta}}\ , \\
\Psi &=  \frac{2 e^{4\delta} \left(\pm 2 a e^{2\delta} (1+e^{4\delta}) + e^{2\gamma} (1+e^{8\delta}) m \right)}{(1+e^{8\delta}) \left(\pm 4 a e^{2\delta} (1+e^{4\delta}) + e^{2\gamma} (1+e^{8\delta}) m \right)}\ , \quad P = \frac{e^{2\gamma} e^{2\delta} m}{4 (1+e^{4\delta})}\ , \\
& S = \frac{\pi e^{2 \gamma} (1+e^{8 \delta}) m \left(\pm 4 a e^{2\delta} (1+e^{4\delta}) + e^{2\gamma} (1+e^{8\delta}) m\right)}{4 e^{4 \delta} (1+e^{4 \delta})^2}\ ,
\end{split}
\eea
where the signs are aligned with the sign in \eqref{eq:bpshor}. We used the subscript in $\beta_\tau$ in order to emphasize that one should think of it as a fictitious (inverse) temperature as it does not relate to a thermal black hole in the standard sense described in the previous section. It can be directly obtained from the BPS limit on \eqref{eq:standardtemp}, but we explain below how to derive it from the resulting smooth Euclidean geometry.

The main insight of \cite{Cabo-Bizet:2018ehj,Cassani:2019mms} in defining the BPS thermodynamics comes in the process of defining the BPS chemical potentials, which use the additional definitions
\be
	\Omega^* := \Omega (a = 0) = 0 , \quad \Phi^* := \Phi (a = 0) = \frac{(e^{8\delta}-1)}{(1+e^{8\delta})} , \quad \Psi^* := \Psi (a = 0) = \frac{2 e^{4\delta}}{(1+e^{8\delta})}\ ,
\ee
i.e.\ the starred quantities are evaluated at the intersection of the extremal and BPS limits. Supersymmetry then ensures the following identity holds,
\be
	M = \Omega^*\, J + 4\, \Phi^*\, Q + 4\, \Psi^*\, P\ ,
\ee
such that the general thermal quantum statistical relation \eqref{eq:qsr} can be now rewritten as
\be
\label{eq:BPSqsr}
	I = - S - \omega\, J - 4\, \varphi\, Q - 4\, \xi\, P\ ,
\ee
where
\be
	\omega:= \beta_\tau\, (\Omega - \Omega^*)\ , \quad \varphi:= \beta_\tau\, (\Phi - \Phi^*)\ , \quad \xi := \beta_\tau\, (\Psi - \Psi^*)\ .
\ee
Note that from the above definition we find the additional identity
\be
\label{eq:omega}
	\omega = \pm 2\, \pi\, {\rm i}\ ,
\ee
as already anticipated, as well as
\be
	\varphi = \pi m\, \frac{e^{2 \gamma} (1-e^{4\delta})}{e^{2 \delta}}  \ , \qquad \xi = - 2 \pi m\, \frac{e^{2 \gamma} e^{2 \delta}}{(1+e^{4 \delta})} \ .
\ee

It is important to stress that, even though they are seemingly defined in a composite way, the chemical potentials $\omega, \varphi, \xi$ are precisely the conjugate quantities to $J, Q, P$ in the BPS limit, as can be checked explicitly by verifying the first law of BPS thermodynamics,
\be
	\delta S + \omega\, \delta J + 4\, \varphi\, \delta Q + 4\, \xi\, \delta P = 0\ .
\ee
This means that the BPS on-shell action $I$ in \eqref{eq:BPSqsr} should be thought of as a function of $\omega, \varphi$ and $\xi$, which are fixed in the grand-canonical ensemble, 
\be
	I (\omega, \varphi, \xi) = \frac{1}{4 \pi}\, \left( \varphi^2 + \xi^2 \right) = \pm \frac{{\rm i}}{2 \omega}\, \left( \varphi^2 + \xi^2 \right)\ ,
\ee
and, as again anticipated, the expression is independent of the fictitious temperature $\beta_\tau$.~\footnote{Note that in the absense of magnetic charges and their conjugate potentials, the on-shell action is formally identical to the one for Kerr-Newman-like black holes in AdS$_4$, \cite{Cassani:2019mms}, except the chemical potentials satisfy a more general constraint than \eqref{eq:omega}. This remains true in the matter-coupled theory and is most easily understood from the gravitational block gluing, \cite{Hristov:2021qsw}, see later.}

Due to the constraint \eqref{eq:omega} it actually follows that one can {\it never} switch the ensemble to fixed angular momentum, but one can for instance look at fixed magnetic charge $P$, 
\be
	\tilde I (\omega, \varphi, P) := I (\omega, \varphi, \tilde \xi(\varphi, P)) + 4\, \tilde \xi (\varphi, P)\, P\ , \quad \frac{\partial I (\omega, \varphi, \xi)}{\partial \xi} \Big|_{\tilde \xi} = 4\, P\ , 
\ee
such that $\tilde \xi = 8 \pi\, P$ and the on-shell action at fixed $\varphi$ and $P$ becomes 
\be
	\tilde I (\omega, \varphi, P) =  \frac{1}{4 \pi}\, \left( \varphi^2 -64 \pi^2\, P^2 \right) = \pm \frac{{\rm i}}{2 \omega}\, \left( \varphi^2 - 64 \pi^2\, P^2 \right)\ .
\ee
We therefore find the mixed-ensemble quantum statistical relation
\be
	\tilde I = -S - \omega\, J - 4\, \varphi\, Q\ .
\ee
It is not a coincidence that in the prepotential language, making use of \eqref{eq:minprep}, the on-shell action in the mixed ensemble can be written as
\be
	\tilde I = \frac{\pi}{2}\, {\rm Im} F(4\, P + \frac{{\rm i}}{2 \pi}\, \varphi)\ .
\ee
It is curious to note that for non-vanishing angular momentum, the on-shell action is not exactly the Legendre transform (with respect to the electric charge) of the entropy since the additional term $\omega\, J$ is non-vanishing and non-removable due to \eqref{eq:omega}. However, the additional identity 
\be
	S^*:= S(a = 0) = S + \omega\, J\ ,
\ee
ensures that the extremal supersymmetric black hole's entropy is still given by
\be
	S^* = -\tilde I - 4\, \varphi\, Q\ . 
\ee

\subsection{Smooth Euclidean saddles}
\label{subsec:wick}

Let us now consider the background solution corresponding to the BPS limit on the parameters discussed above, \eqref{eq:limitmin}-\eqref{eq:alimit} with $\epsilon \rightarrow 0$, where we additionally perform Wick rotation to Euclidean space,
\be
	t \rightarrow {\rm i}\, \tau\ .
\ee
Performing first the usual change of variable, $u = n + a\, \cos \theta$, the BPS limit on \eqref{eq:minmetric} produces the following form of the metric,
\be
\label{eq:BPSmetric}
	{\rm d} s^2 = \frac{\rho^2}{\tilde W (r, \theta)} ({\rm d} \tau+\frac{a\, \tilde L(r) \sin^2 \theta}{\rho^2}\,  {\rm d} \phi)^2 +\tilde W (r, \theta) \left( \frac{{\rm d} r^2}{r^2-a^2} + {\rm d} \theta^2 + \frac{(r^2-a^2) \sin^2 \theta}{\rho^2}\, {\rm d} \phi^2 \right)\ ,
\ee
with
\be
	\rho^2 = r^2 - a^2\, \cos^2 \theta\ ,
\ee
and $\tilde W, \tilde L$ the straightforward BPS limits of the functions $W$ and $L$, respectively. The metric is asymptotic to flat space $\mathbb{R}^3 \times S^1$ as $r \rightarrow \infty$ since in this limit $\tilde W \sim r^2$, with $\{\theta, \phi \}$ the standard spherical angles (and we have used the periodicity of $\tau$ to be shown below). The gauge field from \eqref{eq:gaugefield} is in general complex, as allowed in Euclidean signature:~\footnote{The Einstein equations resulting from \eqref{eq:minlag} are quadratic in the field strengths so the gauge fields can take complex values as long as the metric remains real.}
\bea
\begin{split}
	A = \frac{{\rm i}}{\tilde W^2}\, & \left( \rho^2 (Q\, r + \frac{\partial \tilde V}{\partial \delta}) + \tilde L (\frac{\partial \tilde L}{\partial \delta}-{\rm i}\, a\, P\, \cos \theta) \right) ({\rm d} \tau +\frac{a\, \tilde L \sin^2 \theta {\rm d} \phi}{\rho^2} ) \\
	& + \left( P\, \cos \theta + \frac{{\rm i}\, a\, \cos^2 \theta}{\rho^2}\, ({\rm i}\, a\, P\, \cos \theta - \frac{\partial \tilde L}{\partial \delta}) \right)\, {\rm d} \phi\,
\end{split}
\eea
where $\partial \tilde L/\partial \delta$ and $\partial \tilde V/\partial \delta$ denote the respective BPS limits {\it after} the derivative has been evaluated. Asymptotically, the gauge field is simply
\be
	A = P\, \cos \theta\, {\rm d} \phi\ ,
\ee
and its magnetic dual (see \cite{Chow:2014cca}) is asymptotically
\be
	\tilde A = - Q\, \cos \theta\, {\rm d} \phi\ ,
\ee
as expected in order to recover real electromagnetic charges.

In order to then show regularity of the full solution we need to zoom in on the opposite limit, $r \rightarrow \pm (a + \tilde r)$ with $\tilde r$ small. Using the convenient simplifications
\be
	\tilde L (\pm a) = \frac{e^{2 \gamma} (1+e^{8 \delta}) m \left(\pm 4 a e^{2\delta} (1+e^{4\delta}) + e^{2\gamma} (1+e^{8\delta}) m\right)}{4 e^{4 \delta} (1+e^{4 \delta})^2}\ , \quad \tilde W(\pm a) = a^2 \sin^2 \theta + \tilde L(\pm a)\ ,
\ee
we find 
\be
\label{eq:pseudo}
\lim_{r \rightarrow \pm a} {\rm d} s^2 =  \frac{a^2 \sin^2 \theta}{\tilde W (\pm a)} ({\rm d} \tau -\frac{\beta_\tau}{2 \pi}\,  {\rm d} \phi)^2 +\tilde W (\pm a) \left( \frac{{\rm d} \tilde r^2}{2 a \tilde r} + {\rm d} \theta^2 + \frac{2 \tilde r}{a}\, {\rm d} \phi^2 \right)\ ,
\ee
where the inverse temperature is precisely given by the anticipated expression in \eqref{eq:minbpspotentials}. The resulting space is a smooth manifold, which is however {\it not} flat and does not possess constant curvature. It is nevertheless easy to verify that the coordinates $\{ \tilde r, \phi\}$ with the fixed periodicity $\phi \sim \phi + 2 \pi$ span $\mathbb{R}^2$ (the usual polar coordinates are reached by $\tilde r \rightarrow a R^2$). On the other hand, the periodicity of $\phi$ induces the expected periodicity $\tau \sim \tau + \beta_\tau$, finally resulting in a S$^1$ fibration (the $\theta$ angle) over $\mathbb{R}^2 \times_w$S$^1$, a warped generalization of the usual cigar shaped geometry $\mathbb{R}^2 \times$S$^2$ with shrinking $\mathbb{R}^2$. 

Intriguingly, \eqref{eq:pseudo} is very similar with the near-horizon extremal Kerr geometry, \cite{Bardeen:1999px}, except it exhibits $\mathbb{R}^2$ instead of AdS$_2$ symmetry. Such a similarity also suggests a potential relation with the Kerr/CFT correspondence, \cite{Guica:2008mu}. The fact that the near-horizon limit of the BPS solutions produces a regular space upon Wick rotation and periodic identification of $\tau$ was noticed in a similar setting in \cite{Cvetic:2005zi}, where the authors refer to this as a {\it pseudo horizon}. Due to the symmetry enhancement in this pseudo horizon region, we expect a doubling of the supersymmetry based on the corresponding superalgebra, but have not checked this explicitly. Note that in the extremal limit $a \rightarrow 0$, implying $\beta_\tau \rightarrow \infty$, one needs to start again from the general metric of \eqref{eq:BPSmetric}, finding the expected near-horizon geometry (at $r \rightarrow 0$) of (Euclidean) AdS$_2 \times$S$^2$. Due to the symmetry enhancement of AdS$_2$, one also finds a doubling of the supercharges in this case.

\section{BPS thermodynamics in the STU model}
\label{sec:AdS}
We now consider the more general case of the STU model, which allows us to discuss in more detail the logic behind the BPS limit. It is well-known, see e.g.\ \cite{Hristov:2012bk} for an extended discussion, that the $\cN=2$ superalgebra of Minkowski$_4$ is the so-called Poincar\'e superalgebra. As the name suggests, the bosonic subalgebra consistst of the Poincar\'e group of spacetime translations $P_0$ and $P_i, (i=1, 2,3 )$, spatial rotations $J_{ij}$, and Lorentz boosts $K_i$. Additionally, due to $\cN=2$, there is a complex central charge $Z$ that commutes with the rest of the algebra and accommodates for electromagnetic charges. In the case of stationary solutions such as the black holes, we find that $P_0 = M$, $P_i = 0$, with $M$ the mass of the black hole, and the central charge is evaluated asymptotically, \cite{Ceresole:1995ca}
\be
	Z = \lim_{r \rightarrow \infty} e^{K(z, \bar{z})}\, (X^I (z)\, Q_I - F_I (z)\, P^I)
\ee
with $F_I := \partial F/\partial X^I$ and $K(z, \bar z)$ the K\"ahler potential that is also uniquely fixed by the holomorphic prepotential, given by \eqref{eq:sqrt} in the STU model, see again \cite{Andrianopoli:1996cm,Hristov:2012bk}. The remaining conserved Lorentz charges vanish on the black hole solutions, except for $J_{23} = J$ in the rotating case. For the explicit solutions of interest the asymptotic values of the scalars are not independent parameters, leading to the simple answer \cite{Chow:2014cca}
\be
	| Z |^2 = \sum_{I, J= 0}^3 \left( Q_I\, Q_J + P^I\, P^J \right)\ .
\ee

The BPS limit follows from the simple requirement that some of the fermionic charges of the asymptotic superalgebra are preserved, i.e.\ they vanish on the black holes. Since the anti-commutator of two supersymmetries ${\cal Q}^A_\alpha$ in this case can be evaluated to~\footnote{See again \cite{Hristov:2012bk} for the spinorial conventions and the Clifford algebra of $\gamma$-matrices.}
\be
	\{{\cal Q}^A_\alpha, {\cal Q}^B_\beta \} = M\, \delta_{\alpha \beta} \delta^{A B} + {\rm Re}\, Z\, ({\rm i} \gamma^0))_{\alpha \beta} \epsilon^{A B}+{\rm Im}\, Z\, ({\rm i} \gamma^{123})_{\alpha \beta} \epsilon^{AB}\ ,
\ee 
we find that some of the supercharges will annihilate the black hole solution only if the matrix on the right hand side has some zero eigenvalues, resulting in the BPS constraint
\be
\label{eq:stubpscont}
	M^2 = |Z|^2 = \sum_{I, J= 0}^3 \left( Q_I\, Q_J + P^I\, P^J \right)\ ,
\ee
that ensures half of the supercharges (four out of the original eight real) are preserved. Importantly, the angular momentum does not appear on the right hand side of the supersymmetry anti-commutator in the Poincar\'e superalgebra (in contrast to the AdS superalgebra), and therefore $J = a M$ remains unconstrained. 

Furthermore, in analogy to the minimal case above, it can be shown that the extremal surface~\footnote{Note that in the matter-coupled case there exist two different extremal limits. We are discussing here the {\it fast rotating}, or {\it overrotating}, extremal surface of the STU black holes, which is the analog of the unique extremal surface of the Einstein-Maxwell black holes, see \cite{Chow:2014cca}. The {\it slow rotating}, or {\it underrotating}, extremal surface exists only in theories with scalars and does not intersect with the BPS surface anywhere. In theories with modified fermion sector the underrotating branch instead exhibits hidden supersymmetry at the horizon, see \cite{Hristov:2012nu,Gnecchi:2013mja,Hristov:2018spe}.} only intersects the BPS surface when $J=a=0$, while in the general case the BPS constraint violates the extremality bound and produces naked singularities (regularizable in Euclidean signature).

\subsection{Quantum statistical relation}
\label{subsec:qsrstu}
Similarly to the minimal case, the BPS limit can be enforced on the parameters of the solution in the following way, \cite{Chow:2014cca}
\be
\label{eq:limitstu}
	m \rightarrow m\, \epsilon^2\ , \qquad e^{\gamma_I} \rightarrow \frac{e^{\gamma_I}}{\epsilon}\ ,
\ee
and then sending $\epsilon \rightarrow 0$, while keeping the rest of the parameters unchanged.~\footnote{One can again exchange the scaling of $\delta_I$ and $\gamma_I$, see footnote \ref{ft:1}.} It is easy to show from the solution that in this limit the position of the horizon is again
\be
	r_+ = \pm a\ ,
\ee
upon the change
\be
	a \rightarrow {\rm i}\, a\ .
\ee
Using the full nine-parameter solution of \cite{Chow:2014cca} with arbitrary $\delta_I, \gamma_I$, it is in principle straightforward to evaluate the conserved charges and conjugate chemical potentials, once again finding purely imaginary angular velocity $\Omega$ and momentum $J$ and all remaining thermodynamic quantities remaining real. Unfortunately, the explicit expressions become rather long and unwieldy without offering true novel insight as compared to their values in the minimal case, \eqref{eq:minbpspotentialsang}-\eqref{eq:minbpspotentials}. The interested reader is encouraged to open the {\it Mathematica} file attached to the {\it arXiv} submission source, which contains all the quantities and manipulations omitted here.

Proceeding as in the minimal case, we can define the starred quantities at the intersection of extremality and supersymmetry,
\be
	\Omega^*:= \Omega (a=0) = 0\ , \quad \Phi^{I, *} := \Phi (a=0)\ , \quad \Psi_I^* := \Psi_I (a = 0)\ ,
\ee
obeying the identity
\be
	M = \Omega^*\, J+ \Phi^{I, *}\, Q_I + \Psi_I^*\, P^I\ ,
\ee
such that the quantum statistical relation can be written as
\be
	I = -S - \omega\, J- \varphi^I\, Q_I - \xi_I\, P^I\ ,
\ee
in the ensemble of fixed chemical potentials
\be
	\omega := \beta_\tau\, (\Omega - \Omega^*)\ , \quad \varphi^I := \beta_\tau\, (\Phi^I - \Phi^{I, *})\ , \quad \xi_I := \beta_\tau\, (\Psi_I - \Psi_I^*)\ .
\ee
Again, the above definitions are made meaningful by the first law of BPS thermodynamics,~\footnote{See the attached {\it Mathematica} file for details on the derivation. In the most general case this is checked only numerically due to the complexity of the formulae involved, but various analytic limits exist as well, such as the $X^0 X^1$ model.}
\be
	\delta S + \omega\, \delta J + \varphi^I\, \delta Q_I + \xi_I\, \delta P^I = 0\ ,
\ee
guaranteeing that $I = I (\omega, \varphi^I, \xi_I)$ and is independent of the fictitious temperature $\beta_\tau$. We also recover the same constraint for the angular BPS chemical potential,
\be
	\omega = \pm 2\, \pi\, {\rm i}\ .
\ee

In order to present a meaningful expression for the on-shell action, we need to go to the mixed ensemble of fixed magnetic charges,
\be
	\tilde I (\omega, \varphi, P) := I (\omega, \varphi, \tilde \xi (\varphi, P)) + \tilde \xi_I (\varphi, P)\, P^I\ , \quad \frac{\partial I (\omega, \varphi, \xi)}{\partial \xi_I} \Big|_{\tilde \xi} =  P^I\ , 
\ee
in which case we can check that~\footnote{Note that due to the square root in the general prepotential \eqref{eq:sqrt}, there are different parameter branches similar to the situation for AdS black holes in \cite{Hosseini:2019iad} (see again the attached {\it Mathematica} file). We have not pursued a careful analysis of the solution space.}
\be
\label{eq:osvform}
	\tilde I = \frac{\pi}{2}\, {\rm Im} F(4\, P^I + \frac{{\rm i}}{2 \pi}\, \varphi^I)\ ,
\ee
using the STU prepotential \eqref{eq:sqrt}. Just as in the minimal limit, we have the important additional identity for the extremal supersymmetric entropy $S^*$,
\be
	S^* = S + \omega\, J\ ,
\ee
such that
\be
	\tilde I = - S^* - \varphi^I\, Q_I\ . 
\ee
Note that, just as in the minimal case, the BPS solutions can be regularized in Euclidean signature. The main construction remains exactly the same as in section \ref{subsec:wick} but the explicit formulae become more cumbersome due to the number of independent parameters. We therefore opt not to repeat the analysis and move on to discuss more carefully the main result above, \eqref{eq:osvform}.

\subsection{OSV formula and gravitational blocks}
\label{subsec:osv}
Eq.\ \eqref{eq:osvform} above is in fact the well-known (two-derivartive) OSV formula \cite{Ooguri:2004zv},~\footnote{In order to match precisely the conventions of  \cite{Ooguri:2004zv}, we need the following rescalings:
\be
	p^I_\text{there} = 2 \sqrt{2}\, P^I_\text{here}\ , \quad q_{I,\text{there}} = 2 \sqrt{2}\, Q_{I, \text{here}}\ , \quad \phi^I_\text{there} = \frac{1}{2 \sqrt{2}}\, \varphi^I_\text{here}\ .
\ee} proposed to capture the entropy of supersymmetric (and extremal) black holes in the mixed ensemble and playing an important role in microstate counting due to its relation with the topological string. We have thus {\it physically derived} the OSV formula via BPS thermodynamics in the same sense that \cite{Cabo-Bizet:2018ehj,Cassani:2019mms} derived the entropy functions of \cite{Hosseini:2017mds,Hosseini:2018dob,Choi:2018fdc}. Additionally, upon extremization of the on-shell action $\tilde I$ with respect to $\varphi^I$, one also recovers the black hole attractor equations \cite{Ferrara:1995ih,Ferrara:1996dd} fixing the scalars at the horizon in terms of the electromagnetic charges, (in the present conventions)
\be
	X^I \sim 4 P^I + \frac{{\rm i}}{2 \pi}\, \tilde \varphi^I (P, Q)\ ,
\ee
where $\tilde \varphi^I$ denotes the special value that extremizes $\tilde I$. As we found above, the same OSV formula \eqref{eq:osvform} actually holds generally for the on-shell action of all solutions in the BPS limit, not just the intersection between supersymmetric and extremal ones that consists only of static multi-charge generalizations of the Reissner-Norstr\"om black hole. A crucial difference is the fact that the extremization of the on-shell action gives precisely the entropy and predicts correctly the attractor mechanism {\it only} in the static extremal case. In the general case of arbitrary rotating parameter $a$, we find that the scalars on the {\it pseudo} horizon are no longer constants and exhibiting profiles along the $\theta$ coordinate (see \cite{Chow:2014cca} keeping in mind footnote \ref{ft:2}). 

A more recent development is the explicit interpretation of \eqref{eq:osvform} as a fixed-point formula, \cite{Hristov:2021qsw}. More specifically, it was proposed that the OSV formula can be obtained by adding two gravitational blocks by a particular {\it gluing rule}, producing~\footnote{Note a subtlety in the gluing rule proposals of \cite{Hosseini:2019iad,Hristov:2021qsw} - the overall signs are explicitly dependent on the form of the prepotential. In particular, the sign between the gravitational blocks in the {\it cubic} prepotential is different than the one in the {\it square root} prepotential due to the extra factor of ${\rm i}$ between the two. This means the relative minus sign proposed for the gluing of asymptotically flat black holes in the cubic form \cite{Hristov:2021qsw} flips to a plus when applied here. The original OSV notation using ${\rm Im}$ is applicable in all cases but is less suggestive about the origin of the formula as made up of two independent contributions.}
\be
	\tilde I = \pm \frac{\pi^2}{2 \omega}\, \left( F(4\, P^I + \frac{{\rm i}}{2 \pi}\, \varphi^I) + F(4\, P^I - \frac{{\rm i}}{2 \pi}\, \varphi^I) \right)\ ,
\ee
in the present conventions.~\footnote{The conventions in \cite{Hristov:2021qsw} differ mainly in the rescaling of all chemical potentials $\omega, \varphi^I$ by imaginary prefactors.} Although formally the same expression, this form of the on-shell action featuring a constant 
\be
\omega = \pm 2\, \pi\, {\rm i}\ ,
\ee 
is a direct {\it prediction} from the gluing proposal. In the approach of \cite{Hristov:2021qsw} the constant value of $\omega$ was found from the explicit form of the Killing spinors on the supersymmetric extremal solution, and we now discovered an independent derivation via BPS thermodynamics that makes the above result applicable for a much bigger set of gravitational solutions. This provides further evidence that the asymptotically flat black holes can be understood via the general conjecture of \cite{Hristov:2021qsw}.

\section{Discussion}
\label{sec:outro}
We presented the BPS thermodynamics of a family of rotating Euclidean saddles as a limit of the conventional black hole thermodynamics in Minkowski$_4$, bringing better physical understanding to some of the (previously) dark regions of the solution space (Fig.\ \ref{fig}). 

In the extremal limit we showed that the resulting black hole entropy is given by the Legendre transform of the OSV formula, allowing us to relate our construction with the program of microscopic entropy counting. While this was expected in view of similar results for AdS black holes, \cite{Cabo-Bizet:2018ehj,Cassani:2019mms}, it brings up the question about the role of Euclidean solutions. The Euclidean saddles with AdS asymptotics have a natural interpretation of contributions to the gravitational path integral with given boundary conditions, which in turn also govern the dual CFT deformations.  Such contributions are then automatically counted inside the definition of the relevant dual supersymmetric index. There is no obvious reason why the interpretation of such solutions should be different in the asymptotically flat case, as also exemplified by the very similar structure of the BPS thermodynamics. Yet we should point out that in this case, out of extremality, the BPS saddles exhibit $\mathbb{R}^2$ instead of AdS$_2$ symmetry in the pseudo horizon region, and due to the fixed Minkowski asymptotics we can no longer relate them to (a deformation of) a dual conformal field theory. It would be interesting to understand if a similar structure can be found in the (Euclideanized) D-brane systems that describe the strong coupling regime of the black holes, \cite{Strominger:1996sh,Maldacena:1997de}, and consequently to identify a microscopic description.

An even more general question concerns the remaining dark space (on Fig.\ \ref{fig}) of nakedly singular Lorentzian solutions at $J \neq 0$, situated between the extremal and the BPS surfaces. It seems plausible that such solutions can also be regularized in Euclidean signature, extending the smooth saddles beyond the codimension 1 surface that preserves supersymmetry. We have also not explored the possibility that there exist Euclidean saddles without any Lorentzian analog in matter-coupled theories, e.g.\ due to doubling of the scalar degrees of freedom in Euclidean signature. Such Euclidean saddles could for example allow for chemical potentials $\varphi^I$ that are completely independent of the charges $Q_I$, as suggested in \cite{Bobev:2020pjk}. It is however unclear if they would admit a thermodynamic description.

Another important development concerns the interpretation of the on-shell action for asymptotically flat black holes as a fixed-point formula proposed in \cite{Hristov:2021qsw}, similar to the AdS results of \cite{BenettiGenolini:2019jdz,Hosseini:2019iad}. This proposal would be particularly interesting if successfully applied to multi-centered black holes in Minkowski$_4$, which are well-understood only in the microcanonical ensemble \cite{Denef:2000nb,Denef:2007vg}; the analog of the OSV formula in a mixed ensemble is still missing.\footnote{Instead multi-centered black holes in AdS have not been fully constructed yet, but there is some progress in this direction, \cite{Monten:2021som}.} The careful application of the fixed-point proposal should reveal more about the somewhat mysterious nature of multi-centered black holes and intensify the progress in microstate counting in $\cN = 2, 4$ supergravity theories, see \cite{Cardoso:2019avb,Chowdhury:2019mnb} and references therein.

\section*{Acknowledgements}
I am grateful to Rasim Erol Bekir, Seyed Morteza Hosseini, Stefanos Katmadas, Valentin Reys and Chiara Toldo for valuable discussions, and to Katia Zdravkova for the Photoshop skills, and much more. I am supported in part by the Bulgarian NSF grants N28/5 and KP-06-N 38/11.

\bibliographystyle{JHEP}
\bibliography{tdsotm.bib}

\end{document}